\documentclass{PoS}

\usepackage{subfigure}
\makeatletter
\newcommand{\figcaption}[1]{\def\@captype{figure}\caption{#1}}
\newcommand{\tblcaption}[1]{\def\@captype{table}\caption{#1}}

\title{Finite density QCD phase transition in the heavy quark region}
\ShortTitle{Finite density QCD phase transition in the heavy quark region}
\author{\speaker{H.~Saito}, S.~Aoki, K.~Kanaya, H. ~Ohno\footnote{Current address: Fakult\"at f\"ur Physik, Universit\"at Bielefeld, D-33615 Bielefeld, Germany}\\
   Graduate School of Pure and Applied Sciences, University of Tsukuba, Tsukuba, Ibaraki 305-8571, Japan\\
    E-mail: \email{saitouh@het.ph.tsukuba.ac.jp}}
\author{S.~Ejiri, Y.~Nakagawa\\
   Graduate School of Science and Technology, Niigata University, Niigata 950-2181, Japan}
\author{T.~Hatsuda\\
   Department of Physics, The University of Tokyo, Tokyo 113-0033, Japan}
\author{T.~Umeda\\
   Graduate School of Education, Hiroshima University, Hiroshima 739-8524, Japan}
\author{(WHOT-QCD collaboration)}
\abstract{
We extend our previous study of the QCD phase structure in the heavy quark region to non-zero chemical potentials. 
To identify the critical point where the first order deconfining transition terminates, we study an effective potential defined by the probability distribution function of the plaquette and the Polyakov loop. 
The reweighting technique is shown to be powerful in evaluating the effective potential in a wide range of the plaquette and Polyakov loop expectation values.
We adopt the cumulant expansion to overcome the sign problem in the calculation of complex phase of the quark determinant. 
We find that the method provides us with an intuitive and powerful way to study the phase structure.
We estimate the location of the critical point at finite chemical potential in the heavy quark region. 
}
\FullConference{The XXVIIII International Symposium on Lattice Field Theory, Lattice2011\\
		July 11-16, 2011\\
		The Village at Squaw Valley, US}
\begin{document}

\section{Introduction}
\label{sec:Intro}

In our previous study on the QCD phase structure at zero chemical potential in the heavy quark region \cite{Saito:2011fs}, we have identified the phase structure through the shape of an effective potential $V_{\rm eff}(P)$ defined by the probability distribution function of the plaquette $P$.
Since the plaquette is equal to the gauge action divided by the gauge coupling parameter $\beta$,
we can precisely evaluate $V_{\rm eff}$ in a wide range of $P$ by combining data at different $\beta$'s  via the reweighting technique.

For the same method to work at non-zero chemical potential and/or small quark masses,
reweightings in the direction of the quark mass and the chemical potential are also mandatory.
Therefore, it is useful to introduce new $V_{\rm eff}$ which depends on variables other than the plaquette. 
A study in a light quark mass region by this method is reported in \cite{nakagawa}.

In this report, we consider the phase structure at non-zero chemical potential in the heavy quark region.
At the leading order of the hopping parameter expansion, the influences of dynamical quarks are given in terms of the Polyakov loop $\Omega$.
Using a new $V_{\rm eff}$ which depends on both $P$ and $\Omega_{\rm R} = {\rm Re}\Omega$,
we study the QCD phase structure at zero and non-zero chemical potential.
We adopt the cumulant expansion method \cite{ejiri1,ejiri2}, in order to control the sign problem due to the fluctuation of the complex phase of the quark determinant at finite chemical potential.
The method with new  $V_{\rm eff}$ works well to study the fate of the first order transitions.
We evaluate the location of the critical point  where the first order deconfining transition in the heavy quark limit turns into a crossover, at zero and non-zero values of the chemical potential.

\section{The method}

In this paper, we employ the standard plaquette gauge action and unimproved Wilson quark action, given by
\begin{eqnarray}
S_g &=& -6 N_{\rm site} \beta \hat{P},
\\
S_q &=& \displaystyle \sum_{f=1}^{N_{\rm f}} \left\{ \sum_n\bar{\psi}_n^{(f)}\psi_n^{(f)} 
-\kappa_f \displaystyle \sum_{n} \bar{\psi}_n^{(f)} \left\{ \sum_{\mu=1}^3\left[ (1-\gamma_{\mu})U_{n,\mu}\psi_{n+\hat{\mu}}^{(f)} +(1+\gamma_{\mu})U_{n-\hat{\mu},\mu}^{\dagger}\psi_{n-\hat{\mu}}^{(f)}\right]  \right. \right.  
\nonumber  \\
        & & \left. \left. + e^{\mu_fa}(1-\gamma_{4})U_{n,4}\psi_{n+\hat{4}}^{(f)}+e^{-\mu_fa}(1+\gamma_{4})U_{n-\hat{4},4}^{\dagger}\psi_{n-\hat{4}}^{(f)} \right\} \right\}
\nonumber\\
    & \equiv & \displaystyle \sum_{f=1}^{N_{\rm f}} \left\{ \sum_{n,m} \bar{\psi}_n^{(f)} M_{nm} (\kappa_f,\mu_f) \psi_m^{(f)}\right\}, 
\end{eqnarray} 
where 
$\hat{P}= \frac{1}{18 N_{\rm site}} \displaystyle \sum_{n,\mu < \nu} 
 {\rm Re \ Tr} \left[ U_{n,\mu} U_{n+\hat{\mu},\nu}
U^{\dagger}_{n+\hat{\nu},\mu} U^{\dagger}_{n,\nu} \right]$
is the plaquette, 
$\beta = 6/g^2$ is the gauge coupling parameter,
$\kappa_f$ ($\mu_f$) is the hopping parameter (chemical potential) for the $f$-th flavor, 
and
$N_{\rm site} = N_s^3 \times N_t$ is the lattice volume.
In this study, we mainly consider the degenerate case $\kappa_f=\kappa$ and $\mu_f=\mu$ for $f=1, \cdots, N_{\rm f}$.
Note that $M_{nm}$ we study does not depend on $\beta$.

The probability distribution function for $P$ and $\Omega_{\rm R}  = {\rm Re} \Omega$ is defined by
\begin{eqnarray}
  w(P,\Omega_{\rm R}; \beta, \kappa, \mu) 
  = \int {\cal D} U\, \delta\left(\hat{P}[U]-P\right) \delta\left(\hat{\Omega}_{\rm R}[U]-\Omega_{\rm R}\right) \left[\det M(\kappa, \mu) \right]^{N_{\rm f}}e^{6N_{\rm site} \beta P},
\\
\hat{\Omega}_{\rm R} = {\rm Re} \hat{\Omega}
,\hspace{5mm}
\hat{\Omega} = \frac{1}{3N_s^3} \sum_{\mathbf{n}} {\rm Tr} \left[ 
U_{\mathbf{n},4} U_{\mathbf{n}+\hat{4},4} 
\cdots U_{\mathbf{n}+(N_t -1)\hat{4},4} \right].
\nonumber
\end{eqnarray}

We now define the effective potential as
\begin{equation}
V_{\rm eff}(P, \Omega_{\rm R}; \beta, \kappa, \mu) = -\ln   w(P,\Omega_{\rm R}; \beta, \kappa, \mu) .
\end{equation}
Applying the reweighting technique, we find
\begin{eqnarray}
  V_{\rm eff}(P, \Omega_{\rm R}; \beta, \kappa, \mu) 
  &=& V_0(P, \Omega_{\rm R}; \beta) - \ln \left[ \frac{w(P,\Omega_{\rm R}; \beta, \kappa, \mu)}{w(P,\Omega_{\rm R}; \beta, 0,0)} \right],
\label{eq:Veff}
\end{eqnarray}
where $V_0(P,\Omega_{\rm R}; \beta) = -\ln w(P,\Omega_{\rm R}; \beta, 0, 0)=V_{\rm eff}(P,\Omega_{\rm R}; \beta, 0, 0)$ is the effective potential in the heavy quark limit. 
The second term in the r.h.s.\ is the reweighting factor, which is actually independent of $\beta$ as
\begin{eqnarray}
   \frac{w(P,\Omega_{\rm R}; \beta, \kappa, \mu)}{w(P,\Omega_{\rm R}; \beta, 0,0) }
   &=& \frac{\int {\cal D}U \delta(\hat{P}-P)\delta(\hat{\Omega}_{\rm R}-\Omega_{\rm R}) \left[ \frac{\det M(\kappa, \mu)}{\det M (0, 0)} \right]^{N_{\rm f}}}{\int {\cal D}U\delta(\hat{P}-P) \delta(\hat{\Omega}_R-\Omega_{\rm R})}
   \equiv \left\langle \left[ \frac{ \det M(\kappa, \mu)}{\det M (0, 0)} \right]^{N_{\rm f}}\right\rangle_{\!\!\! P, \Omega_{\rm R}}. \nonumber \\ 
\label{eq:R}
\end{eqnarray}
%
Since the $\beta$-dependence is inherent in $V_0$ only, reweighting in $\beta$ is simply given by
\begin{eqnarray}
V_{\rm eff}(P, \Omega_{\rm R}; \beta, \kappa, \mu) 
= V_{\rm eff}(P, \Omega_{\rm R}; \beta_0, \kappa, \mu) - 6 N_{\rm site} (\beta-\beta_0)\, P.
\label{eq:beta}
\end{eqnarray}

At the lowest order of the hopping parameter expansion, the ratio of the quark determinants in the r.h.s.\ of (\ref{eq:R}) is evaluated as
\begin{equation}
   \frac{\det M(\kappa, \mu)}{\det M(0,0)} 
  =  \exp \left[ 288N_{\rm site}\kappa^4P+3\times2^{N_t+2}N_s^3\kappa^{N_t}\left\{\cosh \left(\frac{\mu}{T}\right) \Omega_{\rm R}+i\sinh \left( \frac{\mu}{T} \right) \Omega_{\rm I}\right\} \right],
  \label{eq:detM}
\end{equation}
where $\Omega_{\rm I} = {\rm Im} \Omega$ is the imaginary part of the Polyakov loop. 
The first term proportional to $P$ can be absorbed into the gauge action by a shift $\beta \rightarrow \beta^* = \beta + 48N_{\rm f}\kappa^4$.
In the evaluation of the effective potential at non-zero chemical potential, the most difficult part
is to evaluate the expectation value  of the complex phase at fixed values of $P$ and $\Omega_{\rm R}$,
denoted by
\begin{equation}
\left\langle e^{ i\theta} \right\rangle_{\! P, \Omega_{\rm R}} \quad
\mbox{with} \hspace{3mm}
\theta=3\times 2^{N_t+2}N_{\rm f}N_s^3\lambda\Omega_{\rm I}
,\hspace{3mm} 
\lambda = \kappa^{N_t}\sinh\left( \mu/T \right).
\label{eq:thetaOmegaI}
\end{equation}

\section{Results at $\mu=0$}
\label{sec:Reweight_Deriv}
\begin{figure}[t]
   \begin{minipage}{7cm}
   \includegraphics[width=5.5cm, angle=270, trim=0 0 0 0, clip]{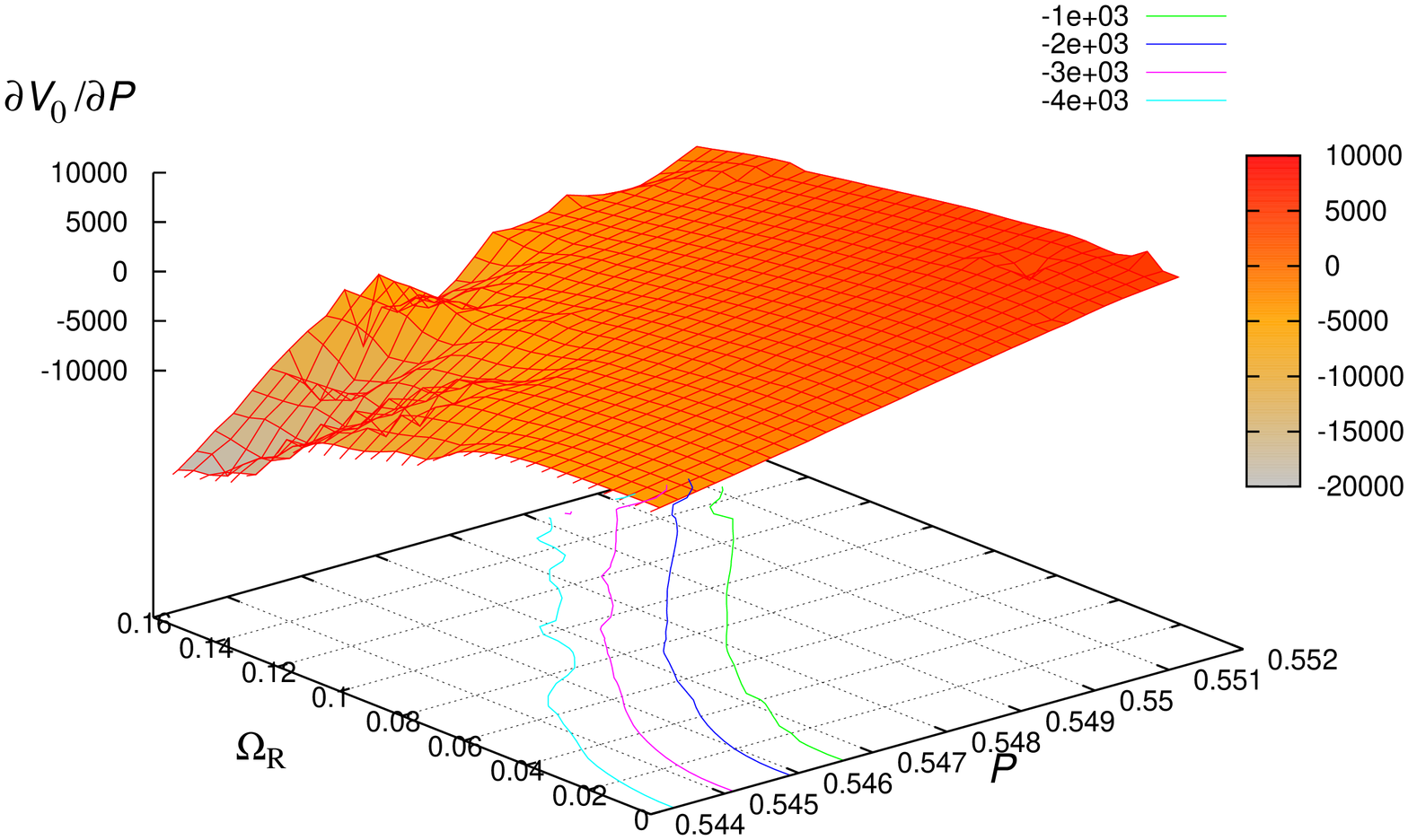} 
   \vspace{-7mm}
   \caption{
   $\partial V_0/\partial P$ at $\beta_0=5.69$.}
   \label{fig:dV0dP}
   \end{minipage}
   \hspace{0.7cm}
   \begin{minipage}{7cm}
   \includegraphics[width=5.5cm, angle=270, trim=0 0 0 0, clip]{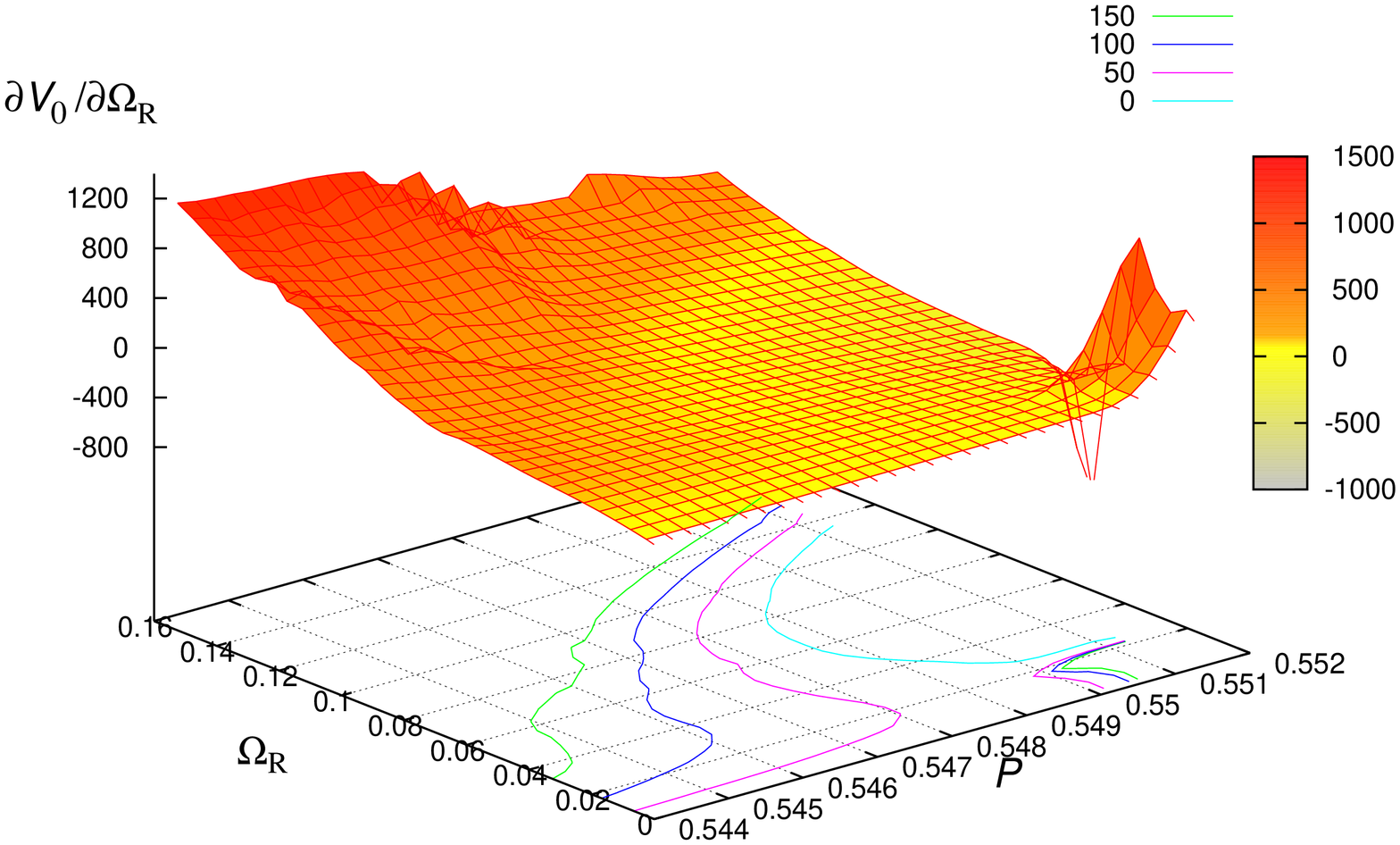} 
   \vspace{-7mm}
   \caption{$\partial V_0/\partial \Omega_{\rm R}$.}
   \label{fig:dV0dOmega}
   \end{minipage}
\end{figure}

Let us first discuss the case of $\mu=0$, in which the imaginary phase term in (\ref{eq:detM}) is absent.
The reweighting formulae for the derivatives of $V_{\rm eff}$ simply become
\begin{eqnarray}
  \frac{\partial V_{\rm eff}}{\partial P}(P, \Omega_{\rm R}; \beta, \kappa)
  &=& \frac{\partial V_0}{\partial P}(P, \Omega_{\rm R}; \beta_0)  - 6N_{\rm site} \left(\beta + 48N_{\rm f}\kappa^4 -\beta_0 \right) ,
  \label{eq:dV0dP}
\\
  \frac{\partial V_{\rm eff}}{\partial \Omega_{\rm R}}(P, \Omega_{\rm R};\kappa) 
  &=& \frac{\partial V_0}{\partial \Omega_{\rm R}}(P, \Omega_{\rm R})  - 3\times2^{N_t+2}N_{\rm f} N_s^3\kappa^{N_t} ,
  \label{eq:dV0dOmega}
\end{eqnarray}
where the argument $\beta$ in $\partial V_{\rm eff}/\partial \Omega_{\rm R}$ and $\partial V_0/\partial \Omega_{\rm R}$ is omitted in (\ref{eq:dV0dOmega}) since they are independent of $\beta$ due to (\ref{eq:beta}).
Note that, besides overall constants [the last terms in (\ref{eq:dV0dP}) and (\ref{eq:dV0dOmega})], 
the dependences of $\partial V_{\rm eff}/\partial P$ and $\partial V_{\rm eff}/\partial \Omega_{\rm R}$ on $P$ and $\Omega_{\rm R}$ are independent of $\beta$ and $\kappa$ [the first terms in (\ref{eq:dV0dP}) and (\ref{eq:dV0dOmega})].

In Figs.~\ref{fig:dV0dP} and \ref{fig:dV0dOmega}, we plot $\partial V_0/\partial P$ and $\partial V_0/\partial \Omega_{\rm R}$ as functions of $(P,\Omega_{\rm R})$. 
Data are taken from the pure gauge configurations generated on a $24^3\times 4$ lattice in \cite{Saito:2011fs}. 
%
To evaluate $w(P, \Omega_{\rm R};\beta,0,0)$, we replace the delta function by the Gaussian function: 
$\delta(x) \approx \frac{1}{\Delta \sqrt{\pi}} \exp \left[ - (x/\Delta)^2 \right]$
where $\Delta =0.0005$ for $P$ and 0.01 for $\Omega_{\rm R}$.
Using (\ref{eq:beta}), we combine data at five $\beta$'s in the range $\beta = 5.68$--5.70 to improve the statistics in a wide range of $P$ and $\Omega_{\rm R}$. 
We also average $\Omega$ over the Z(3) rotation, and remove statistically poor data points with $w < 10^{-135} \times \int \! w\, dP\, d\Omega_{\rm R}$.
We then calculate the derivatives by the difference between $x-\epsilon/2$ and $x+\epsilon/2$,
where $\epsilon = 0.00025$ for $P$ and 0.005 for $\Omega_{\rm R}$.
These parameters are determined by consulting the statistical tolerance of final results. 

\begin{figure}[t]
   \centering
   \includegraphics[width=9.5cm]{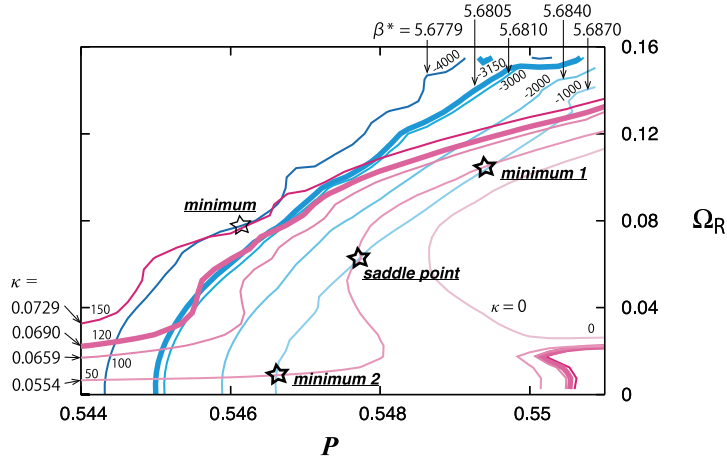} 
   \vspace{-2mm}
   \caption{Contour plot of $\partial V_0/\partial P$ (blue curves) and $\partial V_0/\partial \Omega_{\rm R}$ (red curves) at $\mu=0$. Values of $\beta^* = \beta + 48N_{\rm f}\kappa^4$ and $\kappa$ for the corresponding curves of $\partial V_{\rm eff}/\partial P = 0$ and $\partial V_{\rm eff}/\partial \Omega_{\rm R} = 0$ are also given.}
   \label{fig:cont_mu0}
\end{figure}

To study the phase structure, it is useful to follow the curves $\partial V_{\rm eff}/\partial P = 0$ and $\partial V_{\rm eff}/\partial \Omega_{\rm R} = 0$.
From (\ref{eq:dV0dP}) and (\ref{eq:dV0dOmega}), these curves at different $(\beta,\kappa)$ corresponds to different contour curves of $\partial V_0/\partial P$ and $\partial V_0/\partial \Omega_{\rm R}$, as plotted in Fig.~\ref{fig:cont_mu0}.

When the curves $\partial V_{\rm eff}/\partial P = 0$ and $\partial V_{\rm eff}/\partial \Omega_{\rm R} = 0$ cross at only one point, we have just one minimum of $V_{\rm eff}$.
In this case, the vacuum changes continuously by a small shift of $\beta$ and $\kappa$, indicating no first order transitions around this $(\beta,\kappa)$.
On the other hand, when we have three intersection points (see the contour curves $\partial V_0/\partial P=-1000$ and $\partial V_0/\partial \Omega_{\rm R}=50$ as well as the thick stars in Fig.~\ref{fig:cont_mu0}), we have two minima and one saddle point, implying the existence of the first order transition.
In particular, from the merger of three intersection points, we can determine the critical point where the first order transition line terminates.

From Fig.~\ref{fig:cont_mu0}, we find that three intersection points appear due to the S-shape of the curve $\partial V_{\rm eff}/\partial \Omega_{\rm R}=0$ at small $\kappa$.
The S-shape becomes weaker with increasing $\kappa$, 
and eventually the three intersection points merge to one intersection point at the critical point (e.g.\ the intersection of the contour curves $\partial V_0/\partial P=-4000$ and $\partial V_0/\partial \Omega_{\rm R}=150$ shown by a thin star in Fig.~\ref{fig:cont_mu0} seems to be beyond the critical point).
By consulting the contour curves, we 
find that $\partial V_0/\partial P \approx -3150$ and $\partial V_0/\partial \Omega_{\rm R}\approx 120$ 
(illustrated by bold curves in Fig.~\ref{fig:cont_mu0}) seem to correspond to the critical point.
This gives us a preliminary estimation $\kappa_{\rm cp}\approx 0.0690(7)$ and $\beta^*_{\rm cp}\approx 5.6805(2)$, corresponding to $\beta_{\rm cp}\approx 5.6783(2)$ for the case of $N_{\rm f}=2$, where the errors are estimated from the error of the contour curves due to the statistical errors of potential derivatives.

Our previous study using the effective potential for $P$ gives $\kappa_{\rm cp}=0.0658(3)(^{+4}_{-11})$ and  $\beta^*_{\rm cp}=5.6836(1)(5)$ ($\beta_{\rm cp}=5.6819(1)(5)$) for $N_{\rm f}=2$, where the second errors are  systematic ones due to the method-dependence of the results \cite{Saito:2011fs}.
Our new estimation deviates slightly from the previous one.
We are currently testing a refinement of the method to extract smoother contour curves.
In the remaining part of this report, however, we instead proceed to the case of $\mu\ne0$, by taking an advantage of the relatively simple reweighting formulae for $\mu$.

\section{Results at $\mu \ne 0$}
\label{sec:finite_chem}

At $\mu\ne0$, we need to evaluate the complex phase factor $ \left\langle e^{ i\theta} \right\rangle_{\! P, \Omega_{\rm R}} $ defined by (\ref{eq:thetaOmegaI}).
When $\theta$ fluctuates a lot at large $\mu$, it becomes difficult to estimate $ \left\langle e^{ i\theta} \right\rangle_{\! P, \Omega_{\rm R}} $ reliablly ({\em the sign problem}).

\subsection{The case of phase-quenched finite density QCD}
\label{sec:isospin}

\begin{figure}[t] 
   \begin{minipage}{6cm}
   \includegraphics[width=6cm]{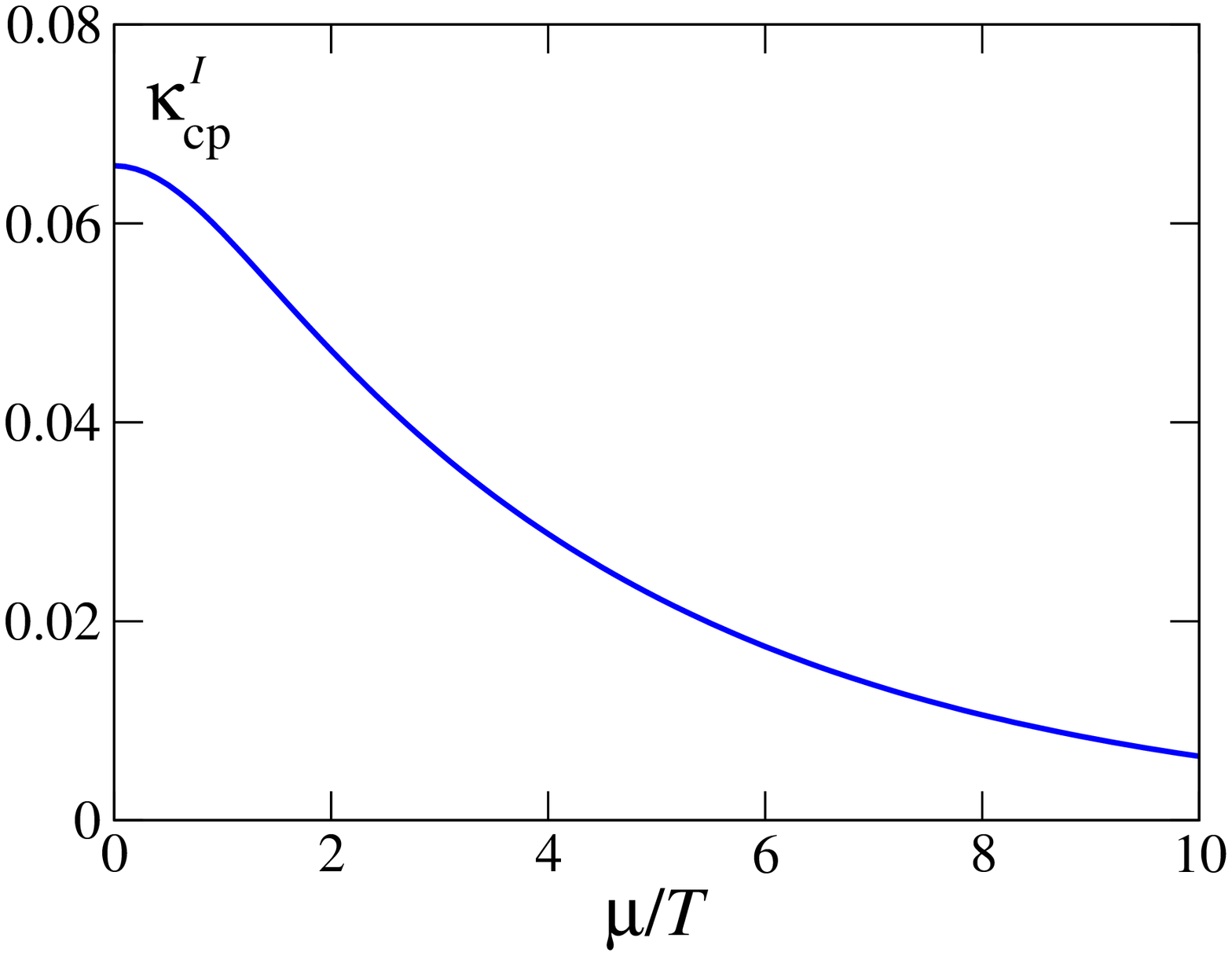} 
   \vspace{-8mm}
   \caption{Critical point in the phase-quenched approximation for $N_{\rm f}=2$.}
   \label{fig:kcp_finitemu}
   \end{minipage}
   \hspace{0.8cm}
   \begin{minipage}{8cm}
   \includegraphics[width=8cm]{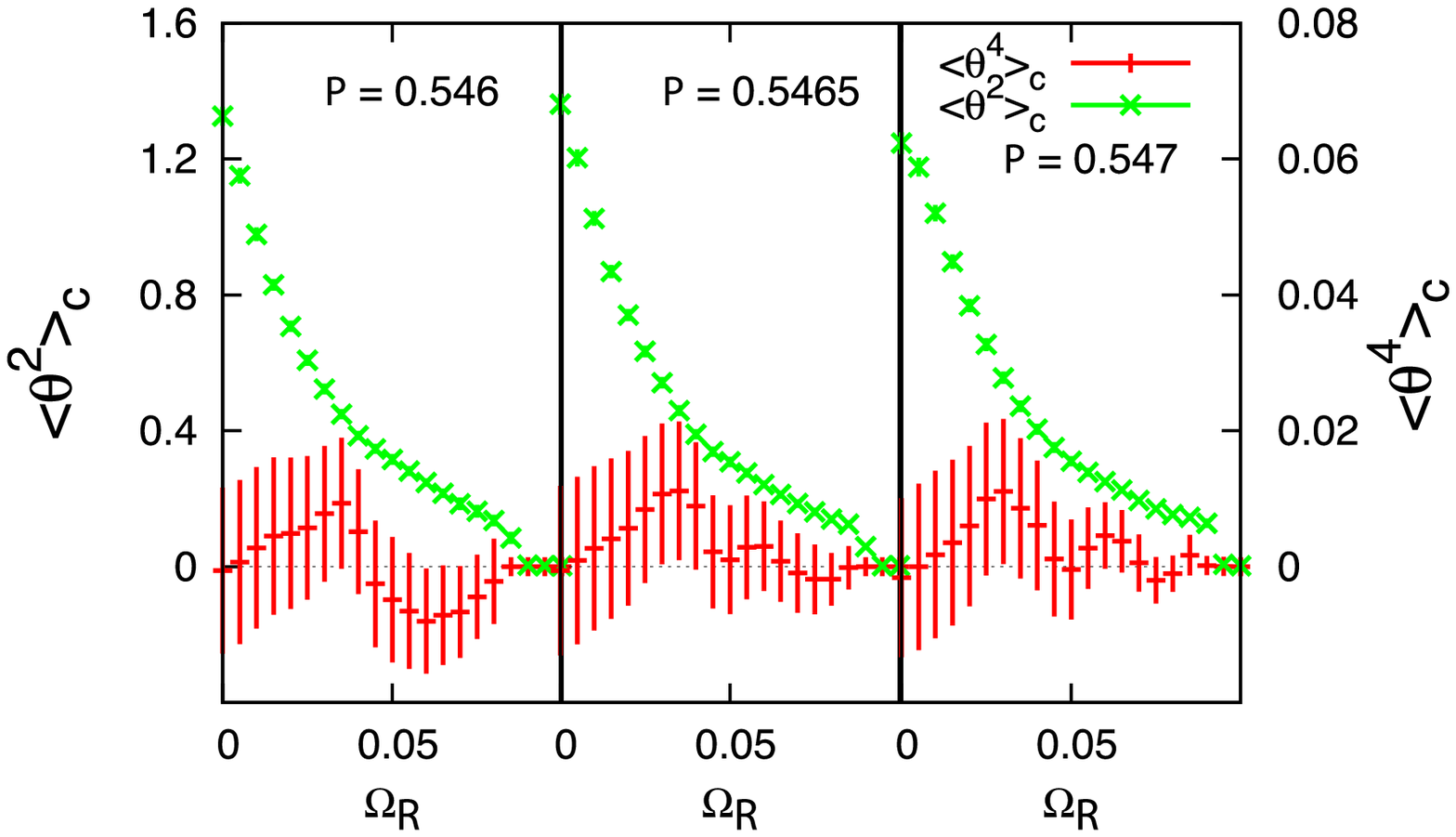}
   \vspace{-8mm}
   \caption{$\left\langle \theta^2\right\rangle_c$ and $\left\langle \theta^4\right\rangle_c$
      at $\lambda=2.0\times 10^{-5}$ 
      around the critical point $(P, \Omega_{\rm R})\approx (0.546, 0.06)$ for $N_{\rm f}=2$.}
      \label{fig:OmegaI24_OmegaR}
   \end{minipage}
\end{figure}

We first consider the case of phase-quenched finite density QCD, in which the complex phase term is removed in the quark determinant ratio (\ref{eq:detM}).
In two-flavor QCD, this corresponds to the case of the isospin chemical potential, $\mu_u=-\mu_d \equiv \mu$. 
After absorbing the first term in the r.h.s.\ of (\ref{eq:detM}) into the gauge action by $\beta \rightarrow \beta^*$, 
we find that the effect of $\mu$ is just to modify the hopping parameter as
$\kappa \rightarrow \kappa\, \cosh^{1/N_t} (\mu/T)$ in the theory at $\mu=0$.
Therefore, to the lowest order of the hopping parameter expansion, the critical point at finite $\mu$ is given by
\begin{equation}
\kappa_{\rm cp}^I(\mu)=\kappa_{\rm cp}(0) / \cosh^{1/N_t}(\mu/T) ,
\label{eq:isospin}
\end{equation}
where $\kappa_{\rm cp}(0)$ is the critical point at vanishing chemical potential. 
See Fig.~\ref{fig:kcp_finitemu}.
Note that, with increasing $\mu$, the critical point approaches towards $\kappa=0$ where the hopping parameter expansion is reliable.

\subsection{Cumulant expansion}

We now turn on the complex phase.
In order to incorporate the effects of the complex phase, we adopt the cumulant expansion method \cite{ejiri1,ejiri2}: 
\begin{eqnarray}
   \left\langle e^{i\theta} \right\rangle_{\! P,\ \Omega_{\rm R}}
   & = & \exp \left[ i\left\langle \theta \right\rangle_c -\frac{1}{2}\left\langle \theta^2 \right\rangle_c - i\frac{1}{3!}\left\langle \theta^3 \right\rangle_c +\cdots \right],
\label{eq:cum}
\end{eqnarray}
where $\theta$ is given by (\ref{eq:thetaOmegaI}), and 
$\left\langle \theta \right\rangle_c \equiv \left\langle \theta \right\rangle$, 
$\left\langle \theta^2 \right\rangle_{c} \equiv \left\langle \theta^2 \right\rangle-\left\langle \theta \right\rangle^2$, 
$\left\langle \theta^3 \right\rangle_{c} \equiv \left\langle \theta^3 \right\rangle -3\left\langle \theta^2 \right\rangle \left\langle \theta\right\rangle +2\left\langle \theta \right\rangle^3$, etc., with the averages taken at fixed $P$ and $\Omega_{\rm R}$.
Since the symmetry of QCD under $\theta \rightarrow -\theta$ implies $\langle \theta^{2n+1} \rangle_c = 0$, the r.h.s.\ of (\ref{eq:cum}) is manifestly real and positive.
This means that, if the cumulant expansion converges, the sign problem is solved.

The distribution of the complex phase in the quark determinant has been shown to be well approximated by a Gaussian distribution up to moderate values of $\mu$ when the complex phase $\theta$ is appropriately defined \cite{nakagawa, ejiri1,ejiri2}.
This means that the leading order term $\left\langle \theta ^2 \right\rangle_c$ dominates in the cumulant expansion, and hence the expansion converges.
In the heavy quark region, we study $\left\langle \theta^{2n} \right\rangle_c
 = \left( 3\times 2^{N_t+2}N_{\rm f} N_s^3  \lambda \right)^{2n}  \left\langle \Omega_{\rm I}^{2n} \right\rangle_c$ with $\lambda=\kappa^{N_t} \sinh( \mu/T )$.
Our results for $\left\langle \theta^2\right\rangle_c$ and the next-leading order $\left\langle \theta^4\right\rangle_c$ are plotted in Fig.~\ref{fig:OmegaI24_OmegaR} for $\lambda=2\times 10^{-5}$.
As will be discussed in the next subsection, this $\lambda$ corresponds to the critical point in the large $\mu$ limit.
Note that the scale for $\left\langle \theta^4\right\rangle_c$ is much magnified in this figure.
We find $\left\langle \theta^2\right\rangle_c \gg \left\langle \theta^4 \right\rangle_c$, in accordance with the Gaussian dominance.
In the followings, we take the leading order approximation of (\ref{eq:cum}) assuming the Gaussian dominance.

\subsection{Critical curve at $\mu\ne0$}

\begin{figure}[t] 
   \includegraphics[width=7.5cm]{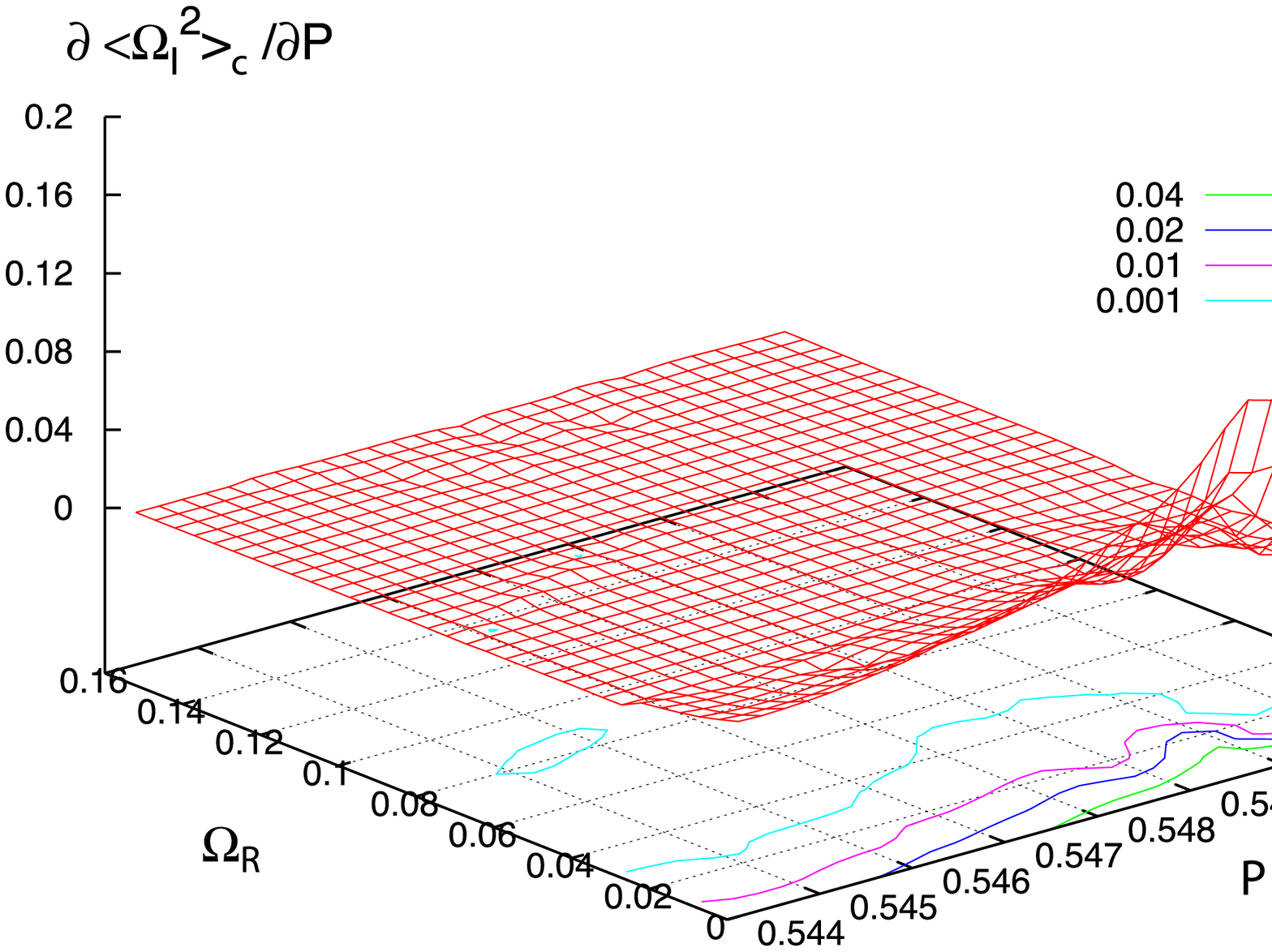} 
   \hspace{0.2cm}
   \includegraphics[width=7.5cm]{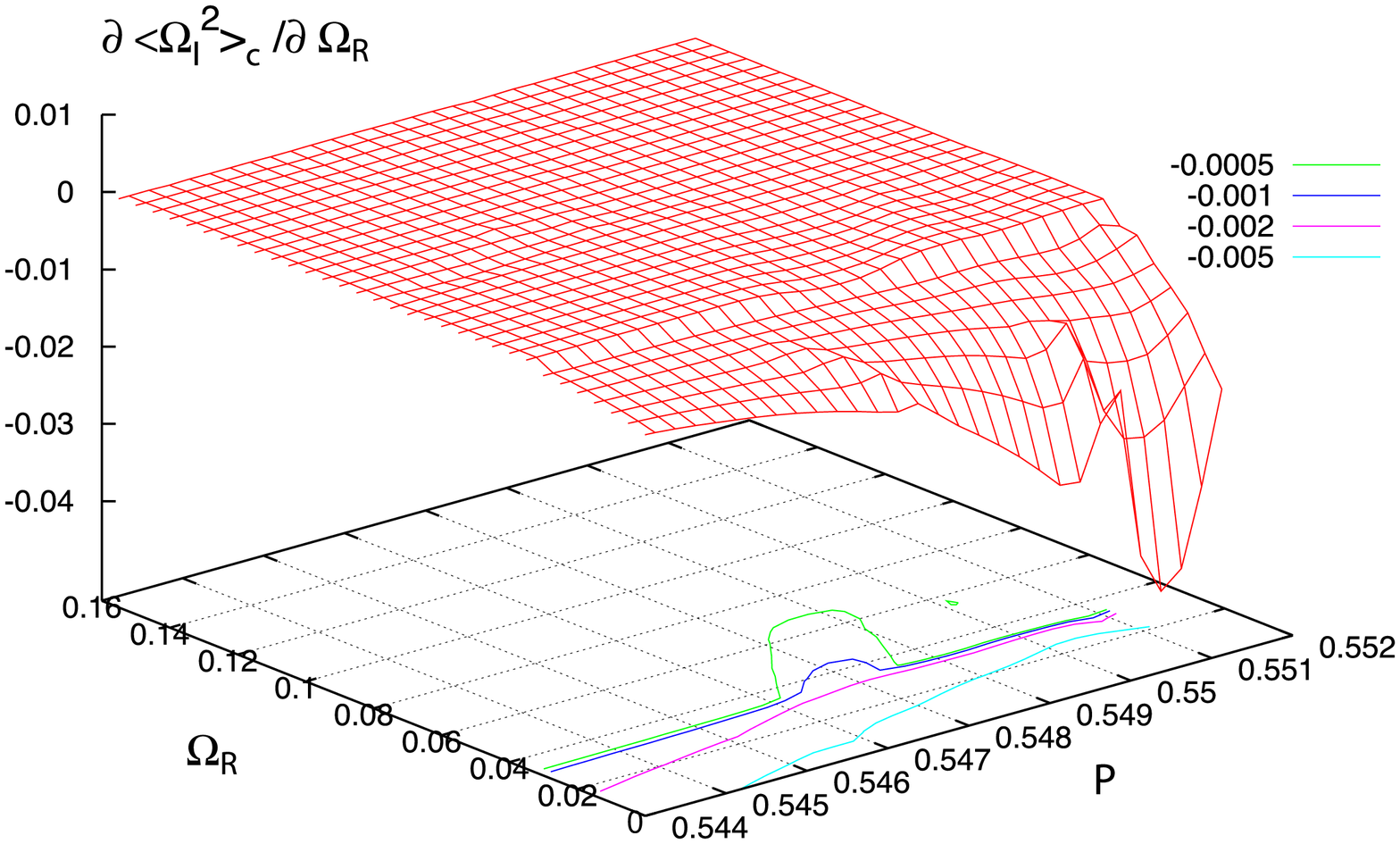} 
   \vspace{-8mm}
   \caption{$\partial \! \left\langle \Omega_{\rm I} ^2\right\rangle_c / \partial P$ (left)
   and $\partial \! \left\langle \Omega_{\rm I} ^2\right\rangle_c / \partial \Omega_{\rm R}$ (right).}
   \label{fig:deriv_OmegaI^2}
\end{figure}

At the leading order of the cumulant expansion, derivatives of $V_{\rm eff}$ read
\begin{eqnarray}
   \frac{\partial V_{\rm eff} }{\partial P} 
   &=& \frac{\partial V_0}{\partial P}-6N_{\rm site} \left(\beta + 48N_{\rm f}\kappa^4 - \beta_0\right)
          +\frac{(3\times 2^{N_t+2}N_{\rm f}N_s^3 \lambda)^2}{2} \frac{\partial \! \left\langle \Omega_{\rm I} ^2\right\rangle_c}{\partial P},
   \label{eq:Veff_mu_P}
\\
   \frac{\partial V_{\rm eff}}{\partial \Omega_{\rm R}} 
   &=& \frac{\partial V_0}{\partial \Omega_{\rm R}}-3\times 2^{N_t+2}N_{\rm f}N_s^3\kappa^{N_t}\cosh \left(\frac{\mu}{T}\right)
           +\frac{(3\times 2^{N_t+2}N_{\rm f}N_s^3 \lambda)^2}{2} \frac{\partial \! \left\langle \Omega_{\rm I} ^2\right\rangle_c}{\partial \Omega_{\rm R}}.
   \label{eq:Veff_mu_Omega}
\end{eqnarray}   
When the last term in (\ref{eq:Veff_mu_Omega}) with the factor $\partial \! \left\langle \Omega_{\rm I} ^2\right\rangle_c / \partial \Omega_{\rm R}$ modifies the S-shaped contours of $\partial V_0 / \partial \Omega_{\rm R}$ shown in Fig.~\ref{fig:dV0dOmega}, the critical point shifts from that of the phase-quenched case given by (\ref{eq:isospin}).
Our results of $\partial \! \left\langle \Omega_{\rm I} ^2\right\rangle_c / \partial P$ and $\partial \! \left\langle \Omega_{\rm I} ^2\right\rangle_c / \partial \Omega_{\rm R}$ are shown in Fig.~\ref{fig:deriv_OmegaI^2}.

We note that  $\partial \! \left\langle \Omega_{\rm I} ^2\right\rangle_c / \partial P$ is quite flat and small around the critical point, and thus just causes a shift of $\beta_{\rm cp}$ according to (\ref{eq:Veff_mu_P}).
We also find that the effect of $\partial \! \left\langle \Omega_{\rm I} ^2\right\rangle_c / \partial \Omega_{\rm R}$ is quite small around the critical point:
When we disregard the last term in (\ref{eq:Veff_mu_Omega}), the critical point locates at $\kappa^{N_t} \cosh (\mu/T) = \kappa_{\rm cp}(0)^{N_t} \approx 2\times10^{-5}$ as discussed in Sect.~\ref{sec:isospin}, i.e.\ $\lambda \approx 2\times10^{-5} \times \tanh (\mu/T)$ along the critical curve.
Using these values and consulting the right panel of Fig.~\ref{fig:deriv_OmegaI^2}, we find that the contribution of the last term in (\ref{eq:Veff_mu_Omega}) is at most about 3\% of the second term around the critical point even in the large $\mu$ limit where $\tanh(\mu/T) \approx 1$.
Therefore, $\kappa_{\rm cp}(\mu)$ is indistinguishable from $\kappa_{\rm cp}^I(\mu)$ shown in Fig.~\ref{fig:kcp_finitemu}, within the current statistical errors.

\section{Conclusion}
\label{sec:conclusion}

We have studied the phase structure of QCD at non-zero chemical potential $\mu$ in the heavy quark region,
using the effective potential defined by the probability distribution function of the plaquette $P$ and the real part of the Polyakov loop $\Omega_{\rm R}$.
The reweighting technique enables us to obtain the effective potential at $\mu\ne0$ in a wide range of $P$ and $\Omega_{\rm R}$.
Adopting the cumulant expansion method to calculate the effects of the complex phase, 
we have shown that the derivatives of the effective potential provide us with an intuitive and powerful way to investigate the fate of first order phase transitions.

We find that the critical point 
where the first order deconfining transition in the heavy quark limit terminates 
locates quite close to that in the phase-quenched case up to large values of $\mu$. 
The smallness of the effects of the complex phase around the critical point is due to the fact that $\kappa_{\rm cp}$ becomes rapidly small as $\mu$ is increased.
Therefore, a careful examination of the effects of the complex phase is required off the critical region and at lighter quark masses.
An attempt to study finite density QCD at light quark masses by combining phase-quenched simulations and the reweighting technique is reported in \cite{nakagawa}.

This work is supported in part 
by Grants-in-Aid of the Japanese Ministry
of Education, Culture, Sports, Science and Technology 
(Nos.~20340047,  
22740168, 
21340049 , 
23540295)  
and by the Large Scale Simulation Program of High Energy Accelerator Research Organization (KEK) Nos.~09/10-25 and 10-09.
SA, SE and TH are also supported in part by the Grant-in-Aid for Scientific Research on Innovative Areas
(Nos.~2004:20105001, 20105003, 2310576). 
HO is supported by the Japan Society for the Promotion of Science for Young Scientists.

\end{document}